\documentclass[prd,nofootinbib]{revtex4}

\usepackage{amsmath,amssymb}
\usepackage{epsfig}

\newcommand{\xgg}{\chi^{\prime\prime}}

\newcommand{\xg}{\chi^{\prime}}

\newcommand{\x}{\chi}
\newcommand{\gc}{\gamma_c}

\def\ba{\begin{eqnarray}}
\def\eqa{\end{eqnarray}}
\def\ea{\end{array}}
\def\beqar{\begin{array}}
\def\bars{\begin{eqnarray*}}
\def\ears{\end{eqnarray*}}

\def\be{\begin{equation}}
\def\ee{\end{equation}}
\def\f{\frac}
\def\g{\gamma}
\def\om{\omega}
\def\ab{\f {\alpha_s N_c}{\pi}}

\def\al{\alpha_s}
\def\l{\label}
\def\th{\theta}

\newcommand{\as}{\ensuremath{\alpha_s}}

\newcommand{\bt}{\begin{tabular}}
\newcommand{\et}{\end{tabular}}
\newcommand{\bd}{\begin{displaymath}}
\newcommand{\ed}{\end{displaymath}\noindent}
\newcommand{\ec}{\end{center}}
\newcommand{\bc}{\begin{center}}

\newcommand{\bi}{\begin{itemize}}
\newcommand{\ei}{\end{itemize}}

\newcommand{\eq}[1]{(\ref{#1})}

\begin{document}
\title{Infrared instability from nonlinear QCD evolution}

\author{R.\ Enberg}
\email{REnberg@lbl.gov}
\affiliation{Theoretical Physics Group, Lawrence Berkeley 
National Laboratory, 
Berkeley, CA 94720, USA}
\author{R.\ Peschanski}
\email{pesch@spht.saclay.cea.fr}
\affiliation{Service de Physique Th{\'e}orique, CEA/Saclay,
  91191 Gif-sur-Yvette Cedex, France\footnote{%
URA 2306, Unit\'e de recherche associ\'ee au CNRS.}}

\begin{abstract}
Using the Balitsky--Kovchegov (BK) equation as an explicit example, 
we show that nonlinear QCD evolution leads to an instability in the propagation 
toward the {\it infrared} of the gluon transverse momentum distribution, 
if one starts with a state with an infrared cut-off. 
This effect takes the mathematical form of rapidly moving 
traveling wave solutions of the BK equation, which we investigate by numerical 
simulations. These traveling wave solutions are different from those governing 
the transition to saturation, which propagate towards the {\it 
ultraviolet}. The infrared wave speed, formally infinite for the leading order 
QCD kernel, is determined by higher order corrections. This mechanism 
could play a r\^ole in the rapid decrease of the mean free path in the 
Color Glass Condensate scenario for heavy ion collisions.  
\end{abstract}

\maketitle

%=======================================================================
\section{Introduction}\label{Introduction}  
%=======================================================================

The properties of perturbative Quantum Chromodynamics (QCD) at high energy 
and/or density presently receive much attention. On phenomenological 
grounds this domain is interesting both for deep inelastic scattering at HERA
and hard scattering at the Tevatron and LHC, and for heavy-ion 
collisions with an initially large density of partons, e.g.\ at RHIC and LHC. 
On theoretical grounds the high energy and density domain is a challenge,
since 
it implies a resummation of logarithms in the perturbative expansion at all orders of 
the strong coupling constant. This resummation goes beyond  the one at the origin of 
the 
linear evolution equation in rapidity $Y$ known as   the 
Balitsky--Fadin--Kuraev--Lipatov (BFKL) equation 
\cite{Lipatov:1976zz}. It implies non-linear terms which describe the damping 
of the scattering amplitudes due to the high density of partons (mainly gluons) 
produced during the evolution. This leads to a transition to {\it saturation} 
\cite{saturation} where a new state of QCD matter, the ``Color Glass 
Condensate'' (CGC) is expected to appear \cite{CGC}. Indeed, the CGC is 
proposed as the initial state formed in heavy ion collisions at high energy 
\cite{maclerran}.

In its mean-field   version, the QCD evolution 
towards saturation is described by the  Balitsky--Kovchegov (BK) nonlinear 
equation  \cite{Balitsky,Kovchegov}. In the color dipole picture of QCD  
\cite{mu94,Nikolaev:1991ja}, it governs the energy dependence of the
onium--target amplitude, where the target is supposed to be interacting 
independently with each dipole resulting from the  evolution of the onium 
wave function with energy. In parton language, and neglecting 
impact parameter dependence, the BK equation describes the 
rapidity $Y$ evolution of the  ``unintegrated gluon distribution'', i.e.\  
the  distribution of gluon transverse momenta $k$ in the target.

An interesting set of results on this  distribution during the transition to 
saturation have already been obtained from the mathematical properties of the 
BK equation. Indeed,   analytic  asymptotic solutions of the 
non-linear equation have been found \cite{Munier} in terms of traveling waves, 
i.e.\  scaling solutions  ${T}(k,Y)\sim {T}\left(k/Q_s(Y)\right),$ where 
$L\equiv \log (k^2/k_0^2)$ and $Y$ play the r\^ole of space and time,
respectively. In fact, the traveling wave pattern of the solutions of the BK 
equation, 
which is already present at 
pre-asymptotic energies \cite{Marquet:2005ic}, 
reflects the earlier proposed \cite{Stasto} ``geometric scaling'' of the $\gamma^* p$ 
cross section.

The saturation scale $Q_s(Y)$  is directly related to the movement in time  of 
the traveling wave and in fact $d\log Q_s(Y)/dY \sim v$ $(v>0)$ is nothing else 
than the speed of the wave which reaches a universal critical 
value at high energies, depending only on the kernel \cite{Munier}.  Hence, in 
physical 
terms 
the gluon transverse momenta propagate with rapidity towards higher and higher 
values, i.e.\  to the {\it ultraviolet}, the typical transverse momentum scale 
being the saturation scale $Q_s.$ This is the key property of the 
transition to saturation. 

In fact the ``final'' result of this evolution---the Color Glass 
Condensate---is supposed to be a fully saturated state where all gluons have a 
momentum equal to (or possibly larger than) the saturation scale characterizing 
the medium. However, in the rapidity evolution from an onium--target scattering,
this ``final'' state is not expected to occur easily at finite energy for the 
whole system. On the other hand, heavy ion collisions at high energies could, due to 
the 
high 
initial density 
of partons, be a good candidate for  the physical occurrence of 
the CGC \cite{maclerran} in the early stages of the collision. However, it is 
known  that difficulties occur in reconciling this appealing 
scenario with the rapid thermalization which seems to occur in the next stages 
of the process. More precisely the mechanisms by which the original CGC, being  
inherently perturbative, feeds the low momentum region 
\cite{Baier:2000sb} are currently under scrutiny \cite{Arnold:2003rq}. Hence, 
this asks the question of the propagation towards the {\it infrared}, when one 
starts with an initial state with a lower transverse momentum limit, or at least 
characterized by a typical high transverse momentum, as is the 
case for the CGC. This propagation has been considered from the point of view 
of a bottom-up mechanism \cite{Baier:2000sb}, possibly helped by plasma 
instabilities \cite{Arnold:2003rq}. However, it is worth considering
the effect of 
energy evolution upon this type of dense initial phase already in the initial stage 
of 
the collision. This is our physical 
motivation for looking at the properties of the BK equation from a new
point of view, the BK equation being considered as a 
prototype for non-linear perturbative QCD evolution. We expect most of our 
results to be valid also for more complete nonlinear evolution equations. 
In the present paper we 
essentially stick to the relevant mathematical properties of the BK evolution, 
leaving a more detailed analysis of the physics of heavy-ion 
collisions for the future.

We will thus discuss how the BK equation gives information on the 
propagation with energy of gluon momenta towards the {\it ultraviolet} and the 
{\it infrared}; namely, how the gluon momentum distribution evolves towards 
larger and smaller momenta, respectively, when starting with  appropriate initial 
conditions. We will consider initial conditions with a  ``sharp'' cut-off, 
either ultraviolet or infrared (see Fig.1), but the mathematical universality 
properties of traveling wave solutions will allow much more general initial 
conditions, provided the initial fronts are steep enough. 
\begin{figure}
\epsfig{file=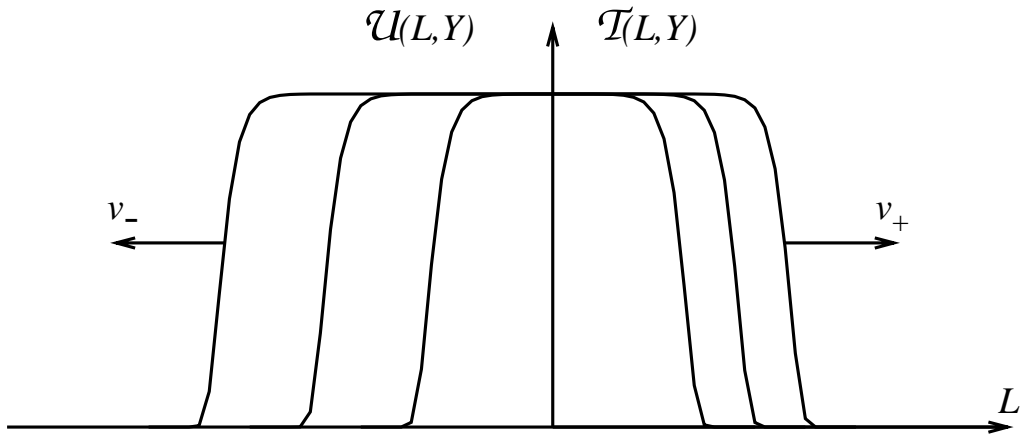,width=15cm}
\caption{\label{fig:fkpp} Sketch of traveling wave solutions for the 
propagation towards the infrared and the ultraviolet. For the 
\emph{ultraviolet} momentum distribution ${\cal T},$ the wave front connecting 
the large-${\cal T}$ and small-${\cal T}$ travels from the left to the right as 
$Y$ increases. For  the \emph{infrared} momentum distribution ${\cal U},$ the 
traveling wave, when it exists (see section \ref{newIV}), travels in the 
opposite direction, from larger to smaller values of the minimum momentum.
The functions ${\cal T},{\cal U}$ are represented for three increasing  
rapidity values.}
\label{travel}
\end{figure}

The plan of our paper is the following. In section \ref{newI} we recall the BK 
equation in transverse momentum space.  In section \ref{newII}, we 
discuss the propagation of momenta towards the ultraviolet (i.e.\  large 
gluon momenta) and show that we recover the same traveling wave solutions and 
scaling properties as  for the solution of the BK equation for the physical 
amplitude.
In section \ref{newIII}, using appropriate initial conditions with an infrared 
cut-off, we derive a new type of scaling solutions of the BK equation for the 
propagation towards the infrared (i.e.\  towards small gluon  momenta). 
In section 
\ref{newIV} we solve the same equations numerically, confirming the analytical 
predictions and study in detail the solutions. Conclusions on the propagation 
properties of the BK equation and its  possible relevance for  the CGC and the 
initial stage of the thermalization process are discussed in section 
\ref{Conclusions}.

%=======================================================================
\section{The Balitsky--Kovchegov equation}\label{newI}
%=======================================================================

The BK equation for the onium--target S-matrix in position space reads
\be
\f{\partial\cal S}{\partial Y}(x_{01},Y) = 
\int d^2x_2\ {\cal K}(x_0,x_1;x_2)\ 
\left\{{\cal S}(x_{02},Y) {\cal S}(x_{12},Y)-{\cal 
S}(x_{01},Y)\right\}\ ,
\l{equation}
\ee
where
\be
{\cal K}(x_0,x_{1};x_{2}) = \ab \ \f {x_{01}^2}{x_{02}^2 x_{12}^2}
\l{kernel}
\ee
is the BFKL kernel \cite{Lipatov:1976zz}. Eq. (\ref{equation})  is the 
impact-parameter 
independent form of the BK equation,  which is what we consider in this work.
$\al$ is
the QCD coupling constant, $x_i$ are the endpoint coordinates of the QCD dipoles 
in 
two-dimensional transverse space, and $x_{ij}=x_i\!-\!x_j$.
To obtain the equation for the amplitude, perform the replacement 
$ {\cal S} \to 1-{\cal N}$ to get
\be
\f{\partial{\cal N}}{\partial Y}(x_{01},Y) = \int d^2x_2\ {\cal 
K}(x_0,x_1;x_2)\ 
\left\{{\cal 
N}(x_{02},Y)\! +\! {\cal N}(x_{12},Y)\!-\!{\cal N}(x_{01},Y)\!-\! {\cal 
N}(x_{02},Y) \, {\cal N}(x_{12},Y)\right\}.
\l{1-equation}
\ee
Again assuming impact-parameter independence, it is well-known \cite{Kovchegov} that  
the Fourier 
transformation to momentum 
space leads to a rather simple  form of the evolution equation which is 
suitable 
 for the derivation of useful analytical properties  \cite{Munier}. Thus one 
introduces the function 
\be
{N}(k,Y) = \int \frac{d^2 x_{01}}{2\pi x_{01}^2} e^{-i  k\cdot 
x_{01}}
{\cal N}(x_{01},Y)
=\int_0^\infty \frac{d r}{r} J_0(kr)\ {\cal N}(r,Y),
\l{r_to_k}
\ee
where $r\equiv |x_{01}|$, and we have integrated over the azimuthal angle. 
Since 
${\cal N}$ only depends on the magnitude $r$ of the vector $x_{01}$, ${ 
N}$ 
depends only  on the gluon transverse momentum magnitude $k$.  ${ N}(k,Y)$ 
can be related to   the unintegrated gluon distribution of the target, and thus 
 gives a direct information on 
 the distribution of transverse momenta of the  gluons probed in the target when 
the total rapidity is $Y.$

The equation \eq{1-equation} then takes the form
\be
\frac{\partial {N}(k,Y)}{\partial Y}
= \ab \left[ K \otimes  {N}(k,Y) -  {N}^2(k,Y) \right]
\l{BK_in_k}
\ee
where the kernel $K$ is the integral kernel of the BFKL equation in momentum 
space. It acts on the gluon distribution as
\be
K \otimes  {N}(k,Y) =
\int_0^\infty \frac{dk'^2}{k'^2}
\left[
\frac{k'^2 {N}(k',Y) - k^2 {N}(k,Y)}{|k^2 - k'^2|} +
\frac{k^2 {N}(k,Y)}{\sqrt {4k'^4 + k^4}}
\right].
\l{integral}
\ee

A comment useful for the further analysis is in order. It 
appears easier to discuss the propagation towards  the {\it ultraviolet} and 
the {\it infrared} using  formulas 
(\ref{BK_in_k},\ref{integral}) in terms of gluon momenta. Indeed, this is 
the formulation for which the mathematical results in terms of traveling waves 
\cite{Munier} are better known. However, the universality properties of the 
traveling wave regime   in momentum space are expected  to be valid 
in 
position space too. Hence, the propagation of momenta towards the ultraviolet 
(infrared) 
 will correspond to the evolution of QCD dipoles towards smaller 
(larger) sizes.

The mathematical argument for this statement is, briefly, that it is easy to 
realize (e.g.\ by means of a  Mellin transform \cite{Lipatov:1976zz,mu94}), that 
the 
linear part of the equation  is symmetric by Fourier transform,  interchanging $k,k'$ 
with  $1/r,1/r'$ in Eq.(\ref{integral}). It is then known from general 
mathematical properties  \cite{ebert,Munier}, that the traveling wave regime does not 
depend of the form of the non-linear term even if  modified by 
Fourier 
transformation.

%=======================================================================
\section{Propagation toward the ultraviolet}\label{newII}
%=======================================================================

Let us first consider the problem of the propagation of the distribution of 
momenta towards the ultraviolet, in an ``universal way'', that is in a way 
independent of the precise physical structure of the target. In order to 
formulate 
the problem, we want to consider  initial conditions defined  only by  a 
sufficiently sharp {\it ultraviolet} cut-off. In fact, interestingly enough, we 
will see that the resulting traveling wave pattern is independent of the initial
cut-off 
profile. 

We start with a simple choice of initial condition, namely a 
$\theta$-function. As we shall see, the asymptotic  
solution is the same for any other initial distribution with a sharp enough 
cut-off\footnote{We check this point numerically in Section \ref{newIV}.}. 
Hence we choose
\be
{\cal T}(k,Y_0) = \th (-L)
\l{boundaryk1}
\ee
where $ L \equiv \log \left({k^2}/{k_0^2}\right),$ with $k_0$ being the initial
ultraviolet 
cut-off. The notation ${\cal T}$ is now for the  solution of the BK equation 
starting from the $\theta$ cut-off (\ref{boundaryk1}). 
When the rapidity increases, the transverse momenta of the gluon distribution  
evolve 
towards larger and larger $k$. As we shall  see, the solution takes 
the form of a traveling wave, the same that was found \cite{Munier} for the 
physical amplitude ${ N}(k,Y)$ itself. The wave front defines a 
critical region in $k$ moving forward with rapidity. 

Let us sketch the arguments in the present case of an initial  $\theta$-function. One 
introduces the Mellin 
transform with respect to the variable $k/k_0.$ 
In the long-range  tail of the distribution, where ${\cal T}(Y) \sim  0$,  
the non-linear 
term in (\ref{BK_in_k}) will remain  small when evolving in $Y$ and can at first be 
neglected. 
The solution of the linear part will thus remain valid. The solution is written
\be 
{\cal 
T}(k,Y) \sim \int_{\cal C} \f{d\g}{2i\pi} \ \f{1}{\g}\ e^{-\g L+  
\om(\g) 
(Y-Y_0)}\ ,
\l{2-equation}
\ee 
where
\begin{equation}
\om(\g) \equiv \ab \chi(\gamma)=\ab \left[2\psi(1)-\psi(\gamma)-\psi(1-\gamma)\right]
\label{chi}
\end{equation}
is the Mellin transform of the kernel\footnote{In the expression 
(\ref{2-equation}) of 
the  Mellin-transform  one may  see  that it is 
symmetric in 
the Fourier transform from momentum to position space, allowing to extend to the 
latter 
the traveling wave properties of the former, as quoted in section   
\ref{newI}.} and the prefactor 
${1}/{\g}$ comes from the initial $\theta$-function condition 
(\ref{boundaryk1}). We are expecting a wave front and thus search for a 
specific velocity $v_+$ characterizing its forward motion. Hence we 
define a comoving frame following  the ultraviolet front and defined by the 
shift  $L\to  L_{WF} \equiv L-v_+ Y.$ A priori, no special velocity $v_+$ is 
selected, 
since for each value of $\gamma$ one has a velocity 
$v_+(\gamma)=\frac{\omega(\gamma)}{\gamma}$ which  stabilizes the 
value of $L_{WF}.$ However, it can be shown from general grounds \cite{ebert} 
that, once the non-linearities are properly taken into account, and provided  
the initial conditions are steep 
enough in $L,$  a critical
velocity
\begin{equation}
v_+=\frac{\omega(\gamma_c)}{\gamma_c}
=\min_\gamma\frac{\omega(\gamma)}{\gamma} \equiv \omega'(\gamma_c)
\label{frontvelocity}
\end{equation}
is selected, in  analogy with the  group velocity in wave physics
\cite{GLR,levin}. Here 
$\gamma_c \approx 0.6275$ is the 
critical anomalous dimension identical to that found for  the BK equation 
\cite{Iancu:2002tr,Munier}. In fact the only condition \cite{Munier} on the 
initial  boundary dependence  is 
${\cal T}(k,Y_0) < e^{-\gamma_c L}.$ This can also be inferred from analyticity 
properties of the integrand of  Eq. (\ref{2-equation}) in 
Mellin 
space  \cite{Munier}, where the  point is to check that the position of the 
prefactor pole  $\f{1}{\g}$  is situated to the left (in $\g$) of the critical 
point $\gamma_c = 0.6275$. This ensures that the overall behavior will be 
driven by the 
critical slope  $\gamma_c$ and not by the one corresponding to the initial 
condition. These conclusions have been tested numerically in Ref.\ \cite{NumBK}.

Indeed, the argument is essentially the same as for the  transition matrix 
element, even if the initial condition is different. Recall that in that case, 
the initial conditions are dictated by the perturbative QCD 
property of ``color transparency'' 
${ 
T}(k,Y) \propto k^{-2}$ at large 
$k$ (i.e.\  color transparency, in terms of dipoles). In mathematical terms, all 
solutions verifying a sharp enough cut-off ${\cal T}(k,Y_0) < e^{-\gamma_c L}$ 
lie in 
the same ``universality class'' and thus follow the same traveling wave pattern 
at high 
enough rapidity. For instance, both the physical amplitude ${\cal N}$ and the 
solution 
${\cal T}$ 
with  
$\theta$-function initial condition lie in this  ``universality class''. 
Hence a 
unifying description of  ``propagation towards the ultraviolet'' is correctly 
defined.
We shall discuss the ``propagation towards the infrared'' within the same 
context in 
the next section \ref{newIII}.

There exists a critical value $k_c(Y)/k_0$ which 
characterizes the position of the moving wave front ${\cal T}$ of momenta 
propagating towards the ultraviolet. The 
value of  $k_c(Y)$ is defined as the value of $k$, at a specified $Y$, where 
the 
distribution ${\cal T}$  has a fixed (and, for consistency, small) constant 
value. Since this
is also the region where these distributions vary significantly with $L,$ see 
Fig.\ \ref{fig:fkpp},   $k_c(Y)$ represents the most probable domain value for 
 momenta as a function of $Y.$ On general grounds, one finds
\begin{equation}
{k_c(Y)} \propto {k_0}\ \exp \left( {v_+ Y 
-\frac{3}{2\gamma_c}\log  Y}\right),
\label{minimal}
\end{equation}
where $\gamma_c$ and  $v_+ $ are given  by the solution of the 
equation (\ref{frontvelocity}). The 
correction $\sim \log  Y$  is a universal retardation effect due to 
the non-linearities. There exists also a third ${\cal 
O}(Y^{-1/2})$ sub-asymptotic universal term in the exponential \cite{Munier} 
which we neglect for our purposes. We also leave for further study the 
interesting 
question of pre-asymptotics and early scaling \cite{Marquet:2005ic,NumBK}.

Together with the front velocity $dk_c(Y)/dY,$ it is possible to obtain some 
general knowledge of the front profile in the region next to the tail, 
including 
information on the diffusive approach to the scaling curve. This means that 
the 
 gluon momenta do not reach instantaneously the scaling behavior of the 
traveling wave. There is a kind of diffusive retardation effect, violating 
geometric scaling. One finds
\begin{equation}
{\cal T}(k,Y) \sim \text{const.}\times
\,\sqrt{\frac{2\pi}{N_c \as \chi ^{\prime\prime}(\gamma_c)}}
\log\left(\frac{k^2}{k_c^2(Y)}\right)
\left(
\frac{k^2}{k_c^2(Y)}\right)^{-\gamma_c}
\exp\left(-\frac{\pi}{2 N_c \as \chi ^{\prime\prime}(\gamma_c)Y}
\log^2\left(\frac{k^2}{k_c^2(Y)}\right)\right)\ .
\label{nfixedalpha}
\end{equation}

A physically  important remark follows from the identification  of the 
traveling wave 
characteristics of general ultraviolet evolution with 
the  amplitude ${\cal N}$ with perturbative initial conditions. It means 
that  
the dynamics are dominated 
by  gluon momenta of order of the saturation scale rather independently of the 
ultraviolet cut-off profile (if sharp enough). In other terms, the propagation 
towards the 
ultraviolet is essentially determined by the evolution of the saturation scale. 
These 
gluons are expected to eventually build the CGC phase.

%=======================================================================
\section{Propagation toward the infrared}\label{newIII}
%=======================================================================

Let us imagine that at $Y=Y_0,$ instead of the boundary condition 
\eq{boundaryk1}, one starts with  boundary conditions for another solution 
${\cal U}$ of the BK equation, that describes 
an initial distribution of  gluons with a   low momentum 
cut-off $k_0$, see\footnote{We keep for convenience the same  value $k_0$ in, 
e.g., 
Fig.~\ref{travel} as for the initial ultraviolet cut-off, but this has no impact on 
the
results.}  Fig.~\ref{travel}.
As in the previous case, we shall discuss an ideal\footnote{In the physical systems 
we have in mind, there should be also a damping factor at $k \to \infty.$ This is required to give a meaningful distribution in position space. Due to the 
universality properties, these limitations are expected to play no important r\^ole 
for our conclusions.} $\theta$-function condition, 
but our 
results will be valid for the whole class of initial conditions lying in the 
same 
``universality class'' of sharp enough infrared cut-offs, to be defined with 
precision 
later on.

We start with the initial condition  
given by
\be
{\cal U}(k,Y_0) = \th (+L),
\l{boundaryk2}
\ee
where ${\cal U}(k,Y)$ denotes the corresponding solution of the BK evolution 
equation.
We will now demonstrate that under very general  conditions such an initial 
condition 
leads to a gluon distribution that propagates rapidly ``backwards'', i.e.\  towards 
the infrared.

However, a special treatment is needed for the BFKL kernel given by 
(\ref{kernel}) in momentum space or (\ref{chi}) in Mellin space. 
The critical value of the speed of the traveling wave is obtained from the 
equation $v=\omega'(\gc)=\omega(\gc)/\gc$, an equation that has one finite and 
one 
infinite solution when $\x$ is the leading logarithmic (LL) BFKL kernel. 
Indeed, $\x$ has one unique finite value of $\gc$  is $\gc=0.6275,$ and one 
formally infinite one corresponding to the pole at $\gamma=0$, see 
Fig.\ \ref{kernelplots}(a). The infinite value (formally for $\gamma=0$) is the 
indication of an 
infrared instability of the propagation which is to be regularized, as will 
become clear later on. The 
analysis of  the asymptotic solutions of  $\cal U$ requires a particular 
treatment in this case. 

\subsection{Diffusive approximation}
Let us proceed in steps and first go back to the original  diffusive 
approximation of the BK equation where  known mathematical properties are useful
to address our problem.
This approximation of the LL BFKL kernel, considered in Ref.\ \cite{Munier}, consists 
in 
expanding 
the kernel around 
$\gamma=\scriptstyle{\f 12}$ and keeping terms up to second order,
\be
\x(\gamma) \simeq \x({\scriptstyle{\f 12}}) + \xg({\scriptstyle{\f 12}}) 
(\gamma-{\scriptstyle{\f 12}}) 
+ \frac{1}{2} \xgg({\scriptstyle{\f 12}}) (\gamma-{\scriptstyle{\f 12}})^2 
\qquad (\text{diffusive approx.}).
\l{diffusive}
\ee
This brings \cite{Munier} the BK equation into the form of the 
Fisher--Kolmogorov--Petrovsky--Piscounov equation \cite{KPP}. In this 
kernel there are no poles and the equation for $\gc$ has two solutions, one 
($\g_+$)
with 
positive, the other ($\g_-$) with negative value $v_-$ of the speed, see 
Fig.\ \ref{kernelplots}(b). The positive value  
yields the same forward traveling waves propagating with speed $v_+$ discussed 
in the previous section.

We now emphasize the significance of the second, negative, value. It will 
correspond to  ``backward'' traveling waves moving with speed $v_-$ towards the 
{\it infrared}.

The existence of  two types of traveling wave solutions for the same equation 
has already been noticed in Refs.~\cite{majumdar}. However, in 
that case they correspond to a statistical physics problem for which both waves
 propagate in the same direction\footnote{We interpret this difference between 
the  statistical physics problem and QCD by the fact that the former is a 
one-dimensional problem analogous to the random breaking of a segment  into 
smaller pieces \cite{majumdar}, while the latter is related to the cascading of 
dipoles in a 
two-dimensional position space (even if the equation under study is its 
one-dimensional projection).} but with 
different speed. In  QCD evolution, the speed $v_-= 
{\x(\g_-)}/{\g_-},$ with $\g_-<0$ and $\x(\g_-)>0$ is negative. As in the 
forward case, there is a universality class of solutions with the same 
pattern, provided the initial condition  for backward propagation is steep
enough in
the backward direction, namely ${\cal U}(k,Y_0) < e^{\g_- L}, L<0.$ This 
universality class is characteristic of the propagation towards the infrared when 
using  the diffusive approximation of the BFKL kernel.

\begin{figure}
\epsfig{file=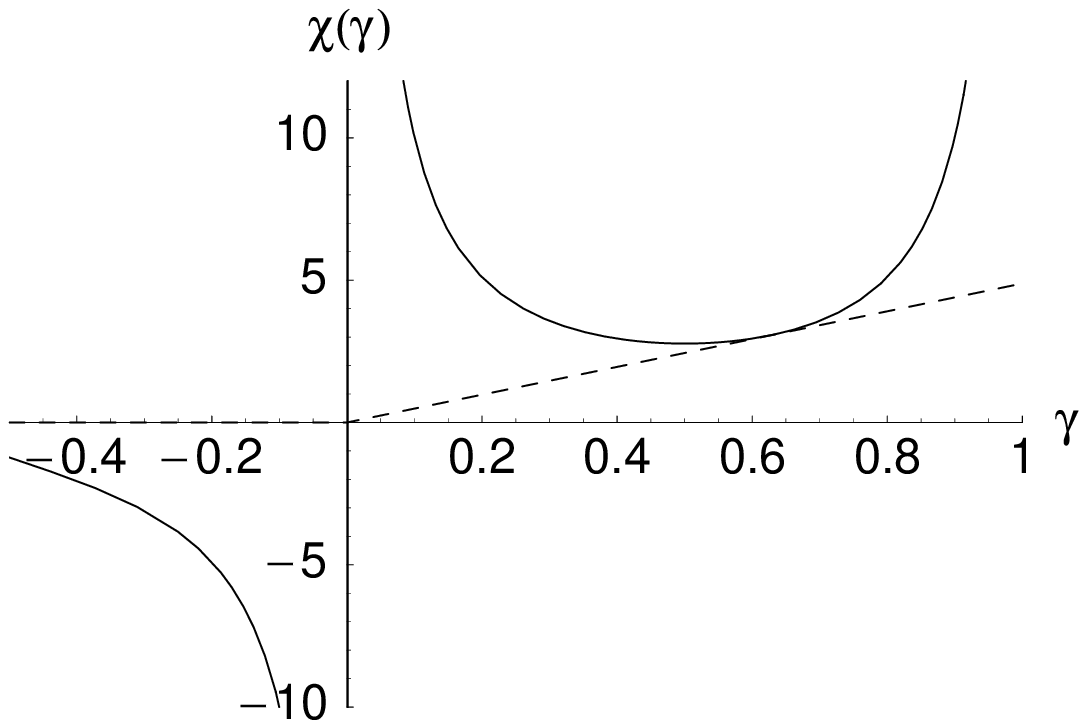,width=0.32\columnwidth}
\epsfig{file=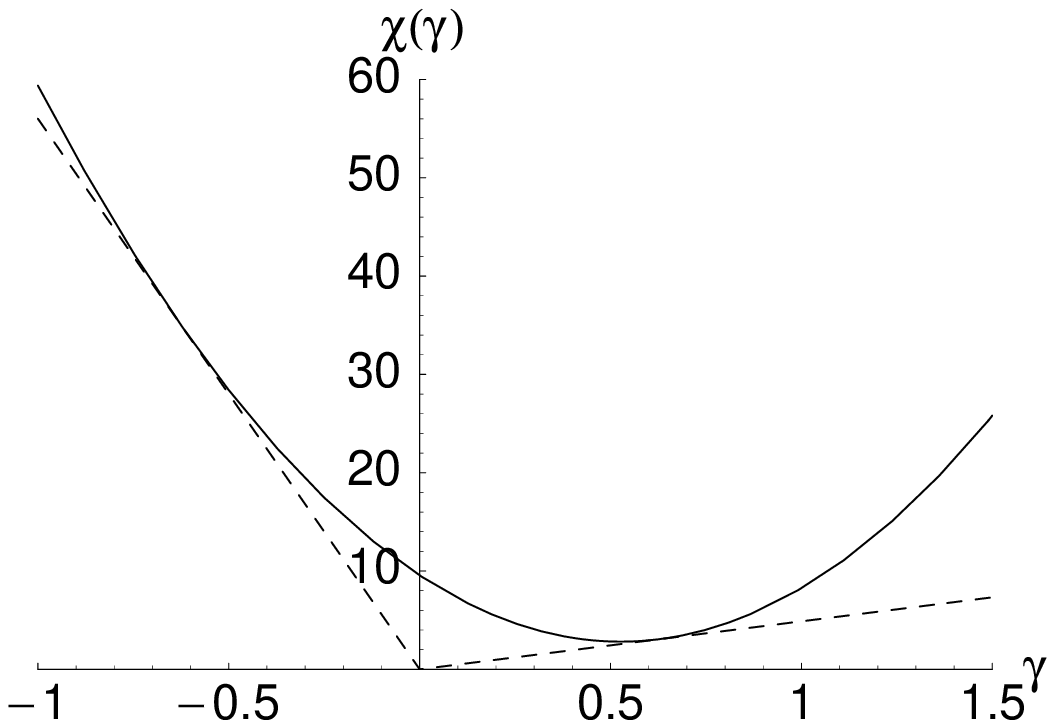,width=0.32\columnwidth}
\epsfig{file=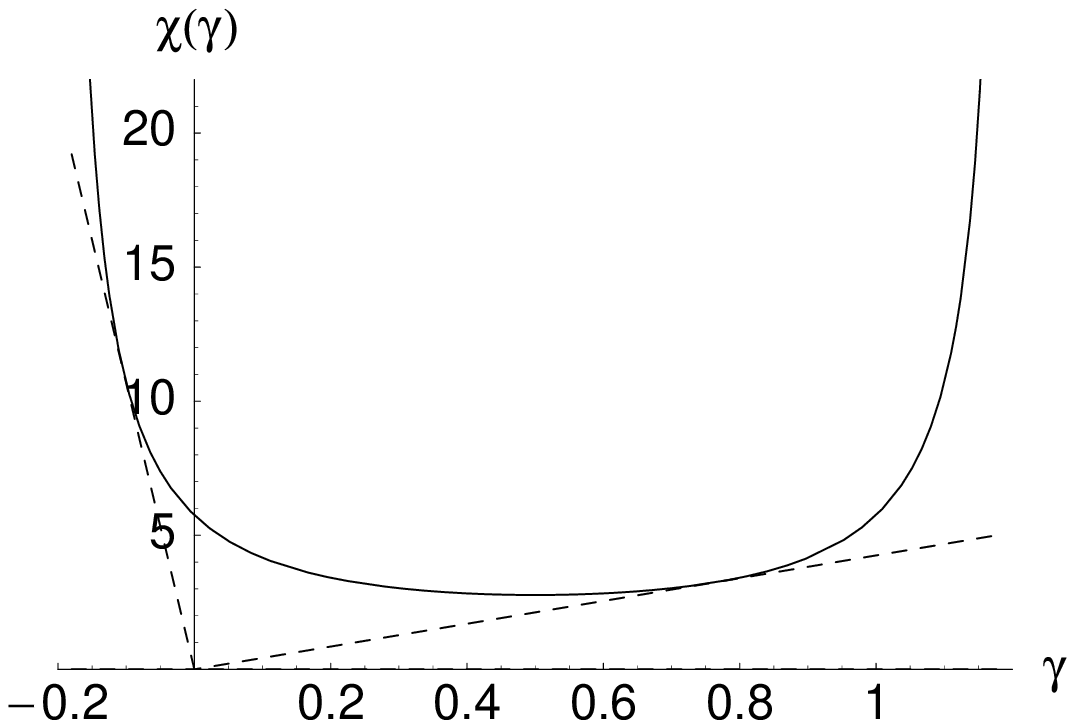,width=0.32\columnwidth}\\
(a) \hspace{5cm} (b) \hspace{5cm} (c) 
\caption{Graphical solutions of the equation $\x(\gc)= \gc \xg(\gc)$, for (a) 
the LL BFKL kernel, Eq.\ \eq{chi}, (b) the diffusive approximation, Eq.\ 
\eq{diffusive}, and 
(c) the two-pole approximation, Eq.\ \eq{twopole} for $\g_0=0.2$. For the 
BFKL kernel there is no solution giving a negative velocity (corresponding to 
negative $\gc$), while for the diffusive and two-pole approximations there 
are both positive and negative velocity solutions. Note that the magnitude of 
the negative velocity (i.e.\ the slope of the tangent) is much larger than 
for the positive velocity (in the LL BFKL case, it is infinite).}
\l{kernelplots}
\end{figure}

\subsection{Regularization by energy-momentum conservation}
We now consider cases where the initial LL BFKL kernel gets modified, see 
Fig.~\ref{kernelplots}(c). This is expected e.g.\ 
from energy momentum conservation constraints and other 
next-to-leading logarithmic (NLL) BFKL effects 
\cite{Salam:1998tj,Peschanski:2004vw}. The pole at $\g =0$ 
gets shifted to a negative value (compare Fig.~\ref{kernelplots}-a and -c). 
Indeed,  the 
effective kernel function $\x(\gamma)$ used in phenomenological calculations 
\cite{Peschanski:2004vw} gets modified along these lines and thus allows for a
finite negative value $\g_-.$ 

A simple argument for understanding the shifting of the pole is  energy 
momentum conservation:\ In 
the relation $\omega = (\alpha_s N_c/\pi) \x(\gamma)$, energy conservation translates 
into 
the condition that $\omega=1$ when $\gamma=0$. In the limit $\gamma\to 0$, 
the BFKL kernel behaves as $\x\sim 1/\gamma$, so we may incorporate energy 
conservation by writing 
\be
\x(\gamma)  \sim  \frac{1}{\gamma + \ab} \qquad (\text{small }\gamma),
\ee
which moves the $\gamma=0$ pole to $\gamma=-\alpha_s N_c/\pi$.

\subsection{Regularization using the  two-pole model}
It is well-known that the LL BFKL kernel is well-approximated by a two-pole 
model, $\x(\gamma)=1/\gamma + 1/(1-\gamma)$, which reproduces the poles of 
the BFKL kernel. This kernel has also been studied in Ref.\ \cite{NumBK} in 
the context of saturation and is found to give results very similar to those
from the full kernel.
 
In view of the above we use as a toy kernel 
\be
\x(\gamma)=\frac{1}{\gamma+\gamma_0}+\frac{1}{1-\gamma+\gamma_0}
+4\left( \log 2 - \frac{1}{1+2\gamma_0} \right),
\l{twopole}
\ee
which reproduces the main features of the BFKL kernel with the poles at 
$\gamma=1$ and  
$\gamma=0$ shifted 
by an amount $\gamma_0$ and $-\gamma_0$ respectively. We consider $\gamma_0$ 
as 
a free 
parameter of our 
model. This kernel can be viewed as an approximation of some next-to-leading 
effects, and is also simple enough to use in our reasoning. It will allow for 
analytical and numerical\footnote{See Ref.\ \protect\cite{NumBK} for the
numerical implementation of the two-pole model.} studies of the infrared propagation 
for
given  $\gamma_0$ 
and its instability when $\gamma_0\to 0.$  The constant term 
is chosen to reproduce the value of the full BFKL kernel at $\gamma=1/2$.

For example, choosing the value $\g_0=0.2$ as illustrated in
Fig.~\ref{kernelplots}(c), one obtains two real solutions to 
the equation defining $\gc$, $\gc^+=0.766$ and $\gc^-=-0.103$, leading to  
traveling wave solutions with both  positive and  negative velocities. All these 
examples, including the $\g_0=0$ case where the infrared instability appears, 
will now be  discussed by comparison 
with numerical simulations.

%=======================================================================
\section{Numerical simulations}\label{newIV}
%=======================================================================

The BK equation has been solved numerically by many groups~\cite{numerical} by direct 
solution of the integro-differential equation. Recently, in Ref.\ 
\cite{NumBK}, it was solved using a discretization of the 
integrand through Chebyshev approximation and a Runge--Kutta method for the 
ensuing system of ordinary differential equation (see \cite{NumBK} for
more details on the method). The solutions were studied in detail and 
compared to analytical results obtained using the traveling wave front 
method.
In this paper we use the same numerical method to study the solutions for the 
distributions $\cal T$ and $\cal U$ in $k$-space. All results that follow are 
obtained 
with a fixed strong coupling constant, $\alpha_s N_c/\pi=0.2$.

First, let us consider the simulation of equation (\ref{BK_in_k}) made using 
the
 original BFKL kernel (\ref{integral}), corresponding to the case of Fig.\ 
\ref{kernelplots}(a).

\begin{figure}
\epsfig{file=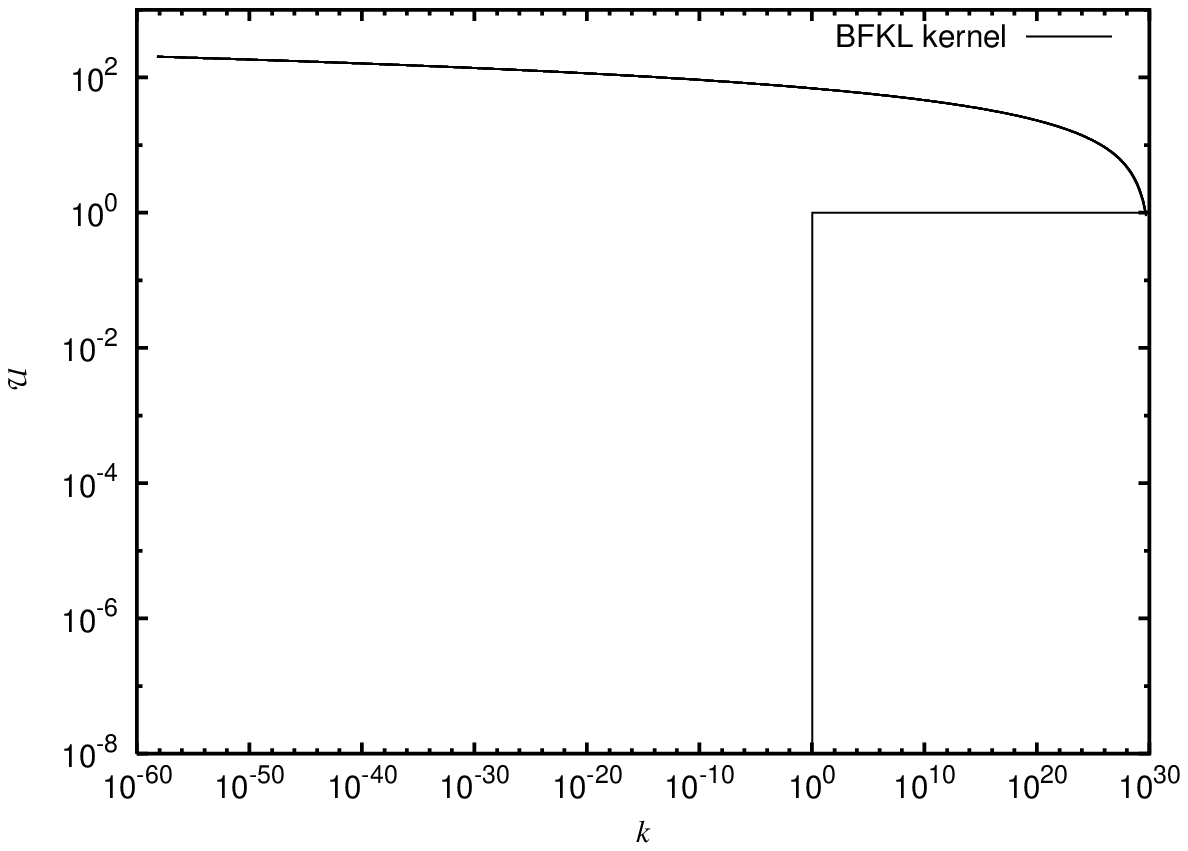,width=0.5\columnwidth}%
\epsfig{file=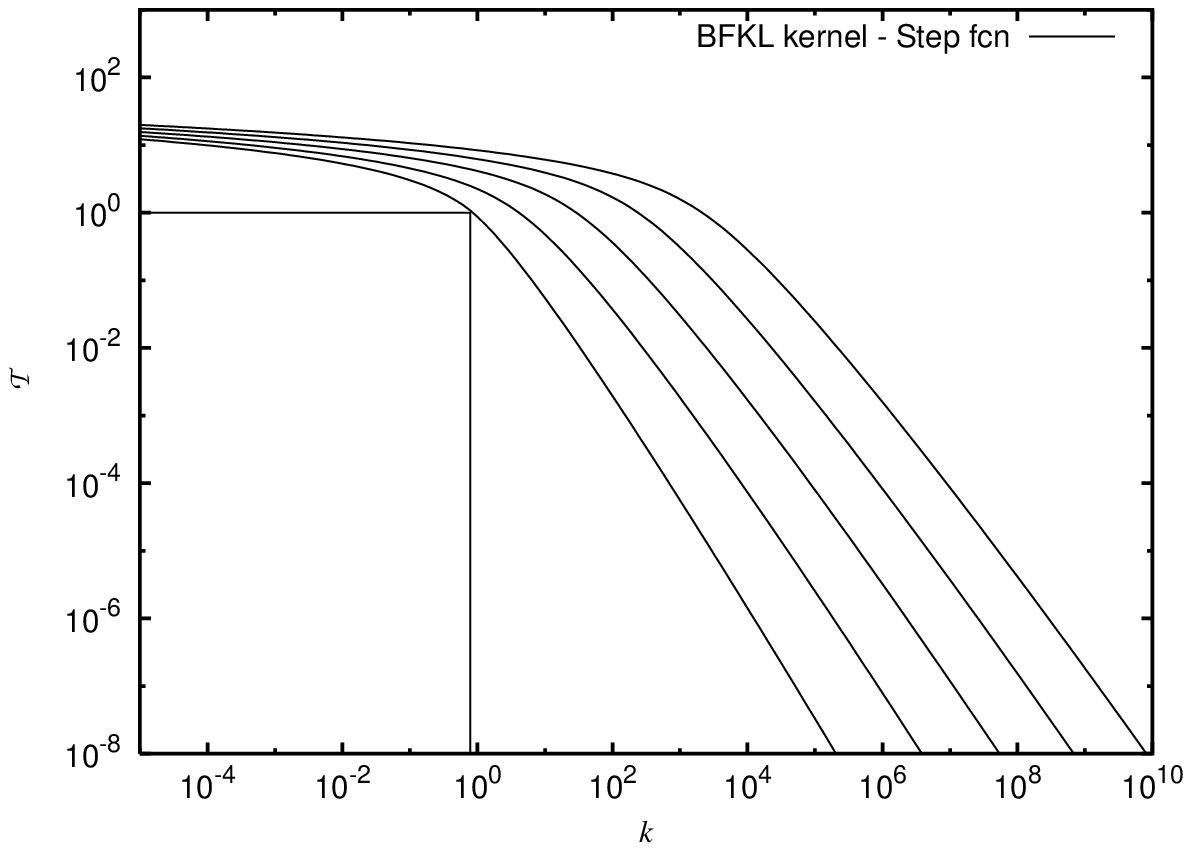,width=0.5\columnwidth}
\caption{Propagation of momenta for the LL BFKL kernel. Left: 
towards the 
infrared; Right: 
towards the ultraviolet. The rapidity steps are $\delta Y = 1.$ In the left 
plot, all lines 
with $Y \ne  0$ scale on the same curve.}
\l{bfklplots}
\end{figure}

We observe that traveling wave solutions are generated in the forward 
direction, while the backward direction is characterized by a fixed (non-traveling) 
scaling curve, obtained {\it almost} immediately in the 
rapidity 
evolution. Moreover, the solution extends also immediately towards very 
negative 
values of $L\equiv \log(k^2/k_0^2),$ i.e.\  in the very far infrared. This is a
clear sign of  infrared instability of the evolution with the LL kernel.

In LL BFKL, hence, only forward traveling waves are present. The backward 
solution  propagates instantly to high $Y$, and generates  a fixed scaling 
solution. In a formal way, the speed gets infinite and the wave front is 
situated at 
$L=-\infty$. This situation is illustrated in Fig.\ \ref{bfklplots}.

In order to analyze in more detail this infrared instability, we considered the 
same 
problem with much smaller steps in rapidity, see Fig.\ \ref{smallsteps}. We see 
the 
evidence for a rapid diffusive evolution towards the scaling curve. Note also 
the  far infrared region, which is immediately reached by the 
evolution. This is a feature related to the singularity of the  canonical BFKL 
kernel at $\g =0,$ and, as we shall see now, is physically tamed by
next-to-leading 
corrections. This sensitivity to next-to-leading QCD  
corrections is a specific and  
interesting feature of backward propagation towards the infrared.

\begin{figure}
\epsfig{file=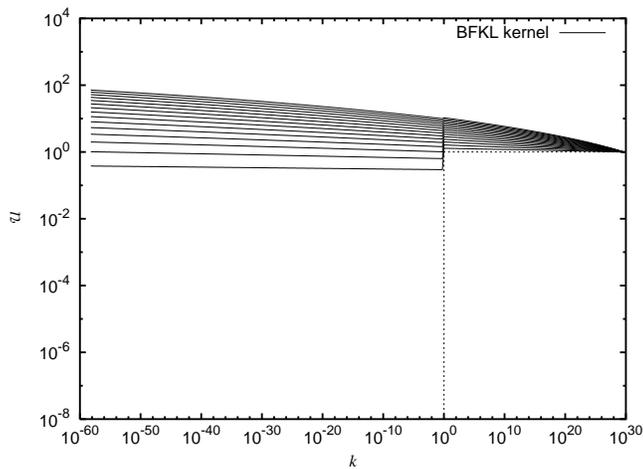,width=0.5\columnwidth}
\caption{First steps of BFKL evolution  towards the infrared. 
The rapidity steps are $\delta Y = 0.1.$}
\l{smallsteps}
\end{figure}

In Fig.\ \ref{twopolefig}, we show the similar plot for the non-linear 
evolution 
using the two-pole kernel  (\ref{twopole}) with different values of the 
parameter 
$\g_0$ which, in our toy model,  schematically  characterize the amount of 
next-leading corrections. Interestingly enough, the  toy model results for 
$\g_0 
=0$ reproduce the same  features as for the BFKL kernel, with a scaling curve 
in 
the backward propagation. For non-zero  $\g_0 =0.2,0.4$, left-moving traveling 
waves 
appear, in agreement with the analytical predictions. In all cases the speed $v_-$
is 
larger than for the forward propagation (i.e.\ larger than the ordinary forward 
BK 
traveling 
waves). Hence 
the infrared instability is made finite but the propagation towards the 
infrared is still much more rapid. In terms of QCD dipoles, one may say that   
dipoles with large sizes are easily produced during  the energy 
evolution, in the ``universality class'' of backward propagation.
\begin{figure}
\epsfig{file=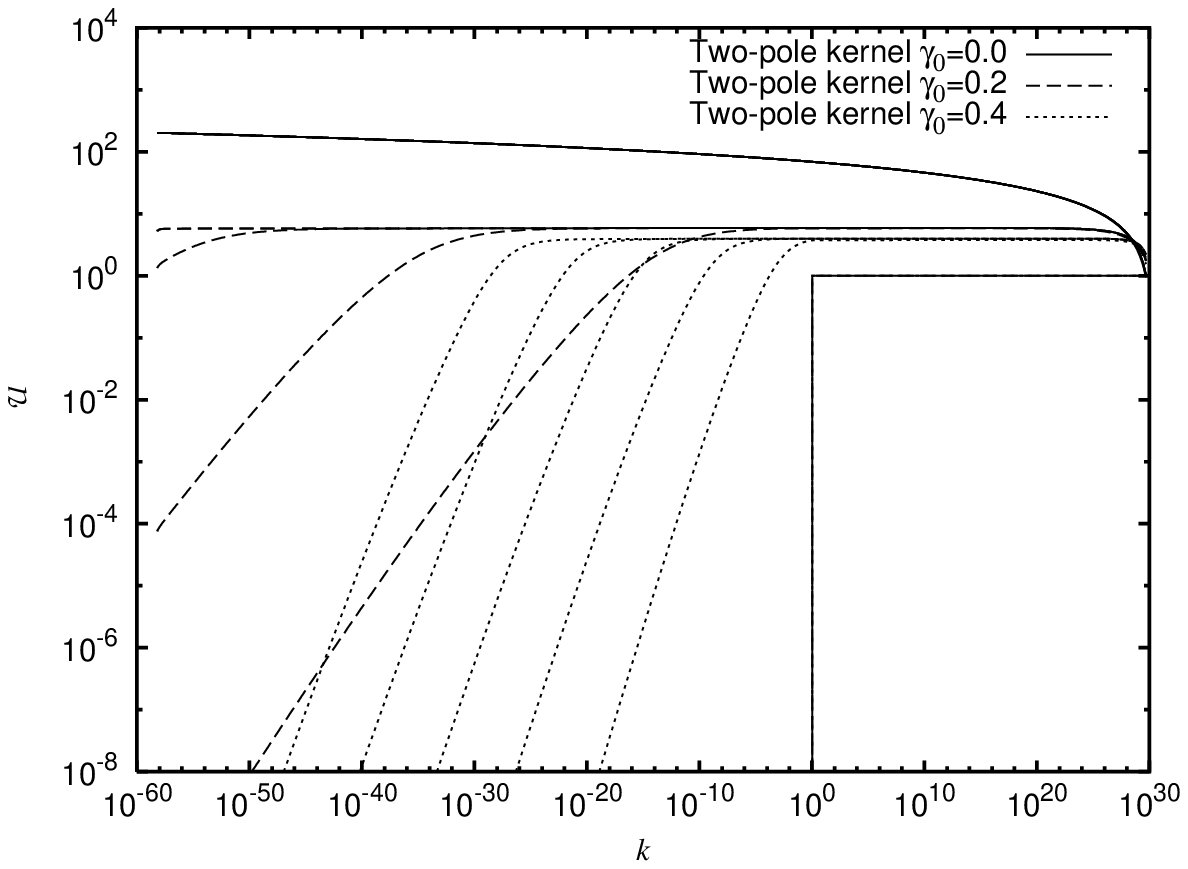,width=0.5\columnwidth}%
\epsfig{file=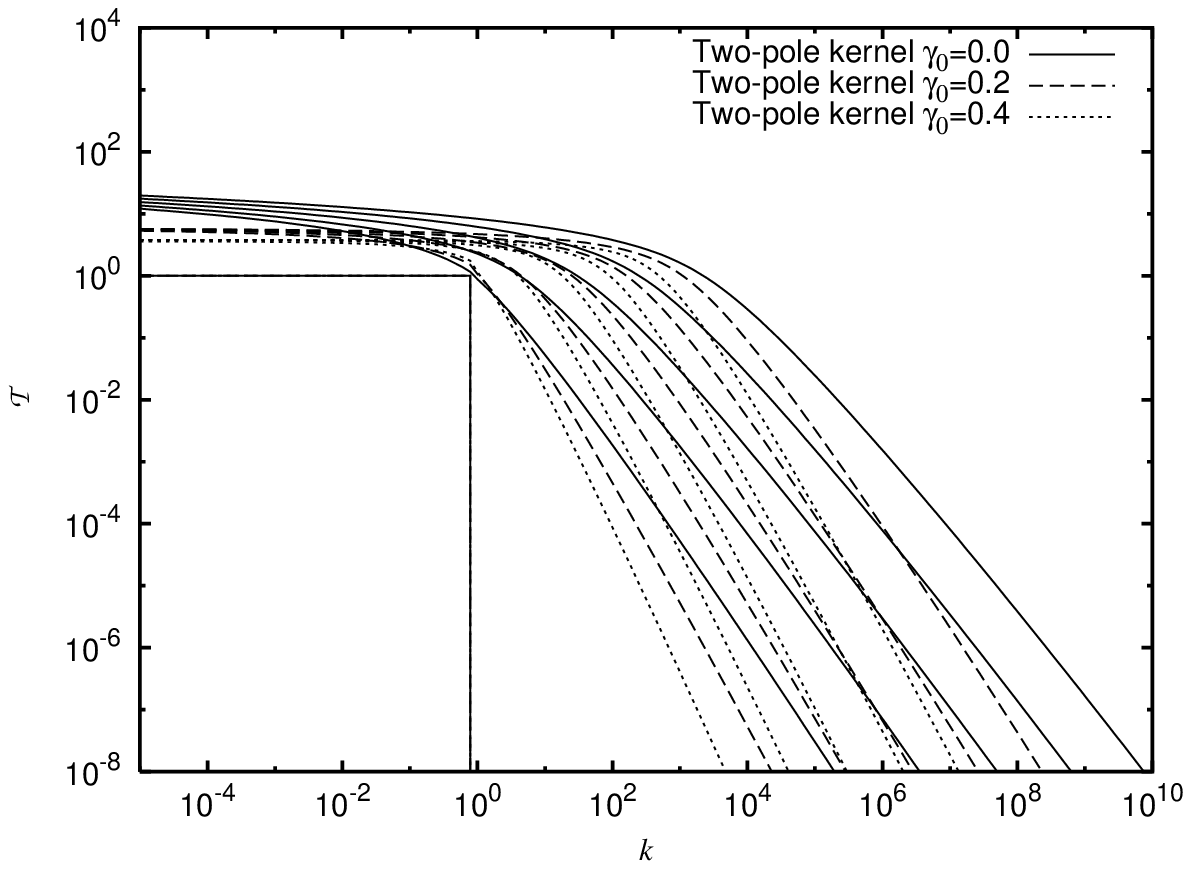,width=0.5\columnwidth}
\caption{Propagation of  momenta for the two-pole model. Left: 
towards the infrared; Right: towards the ultraviolet. The curves are drawn for 
values $\g_0= 0,0.2,0.4$ and the rapidity steps are $\delta Y = 1.$}
\l{twopolefig}
\end{figure}
The comparison between the forward and backward evolutions is exemplified  in 
Fig.\ 
\ref{twopole.4fig}.

\begin{figure}
\epsfig{file=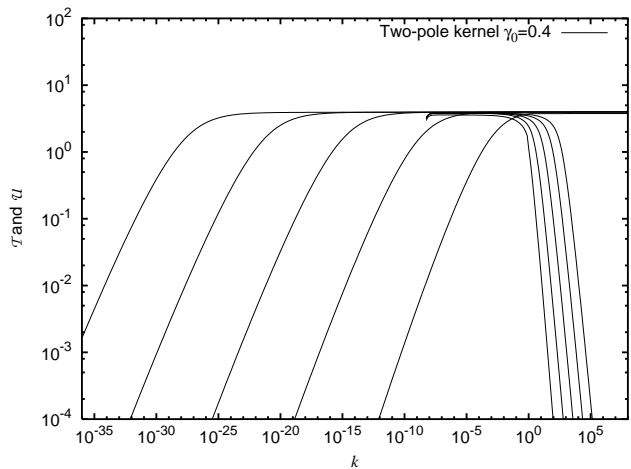,width=0.5\columnwidth}
\caption{Comparison of forward and backward propagation for $\g_0= 0.4$. The 
rapidity 
steps are $\delta Y = 1$}
\l{twopole.4fig}
\end{figure}

To round up the numerical results we give an example of the (in)dependence of 
the traveling 
wave regime from the precise shape of the cut-offs in the 
initial conditions provided they are sharp enough, either in the forward or 
the 
backward direction respectively, to satisfy the mathematical requirements for 
creating traveling waves. We thus  compare initial step functions with an initial 
form \cite{maclerran} of the cut-off profile which is   smoother\footnote{In practice 
we used the numerical Fourier transform 
from $r$-space
of~\protect\cite{maclerran}
\begin{equation}
1 - \exp\left(-r^2 Q_s^2 (Y) / 4  \log[e+1/(r^2 \Lambda^2)]\right)\ .
\end{equation}}
than a $\theta$-function, while satisfying the  sharpness condition. As typical 
examples (see 
Fig.\ \ref{twopoleBFKLfig}) we considered the backward waves obtained  for the 
toy 
model and the forward waves for the original BFKL kernel. 
\begin{figure}
\epsfig{file=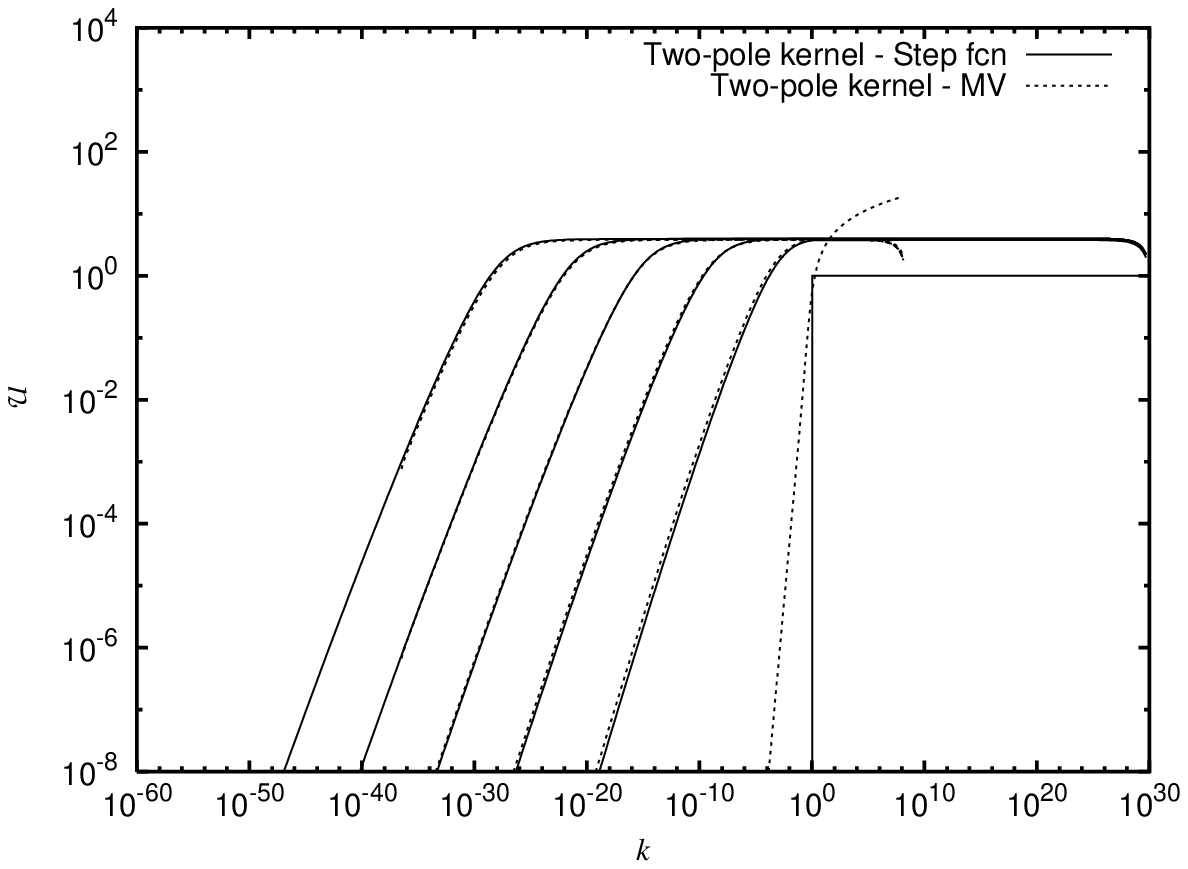,width=0.5\columnwidth}%
\epsfig{file=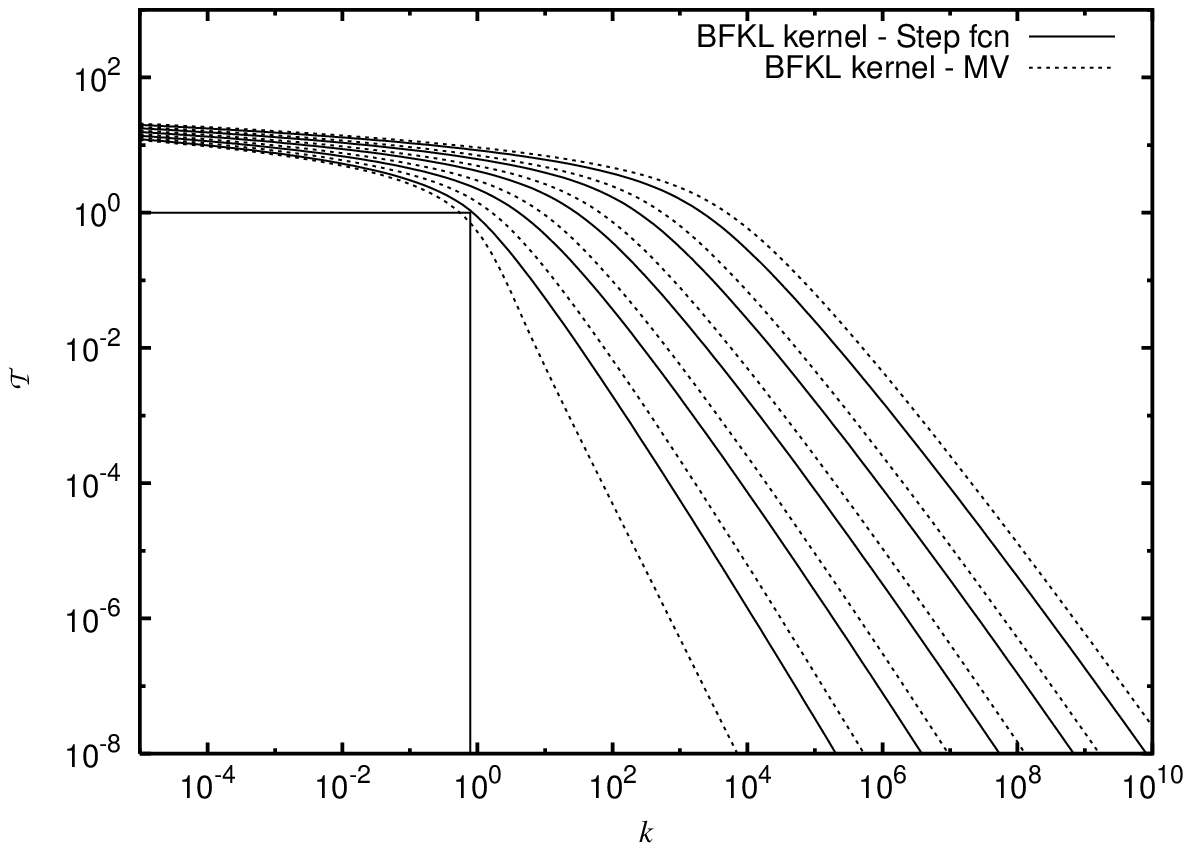,width=0.5\columnwidth}
\caption{Propagation  of  momenta  for different initial conditions. 
Left: two-pole model towards the infrared; Right: BFKL kernel towards the 
ultraviolet. Continuous lines: initial step functions. Dotted lines: 
initial smoother cut-off (see text). The curves are drawn for the BFKL kernel 
and the 
rapidity steps are $\delta Y = 1.$}
\l{twopoleBFKLfig}
\end{figure}

Finally, we confirm that the speed of the front propagation is correctly given 
by the 
analytical formula for asymptotically large rapidities, i.e., we compare the 
rapidity 
dependence of the saturation scale of the forward and backward traveling waves 
obtained from 
the numerical solution to the analytical formula for asymptotically large 
rapidities. 

\begin{figure}[th]
\epsfig{file=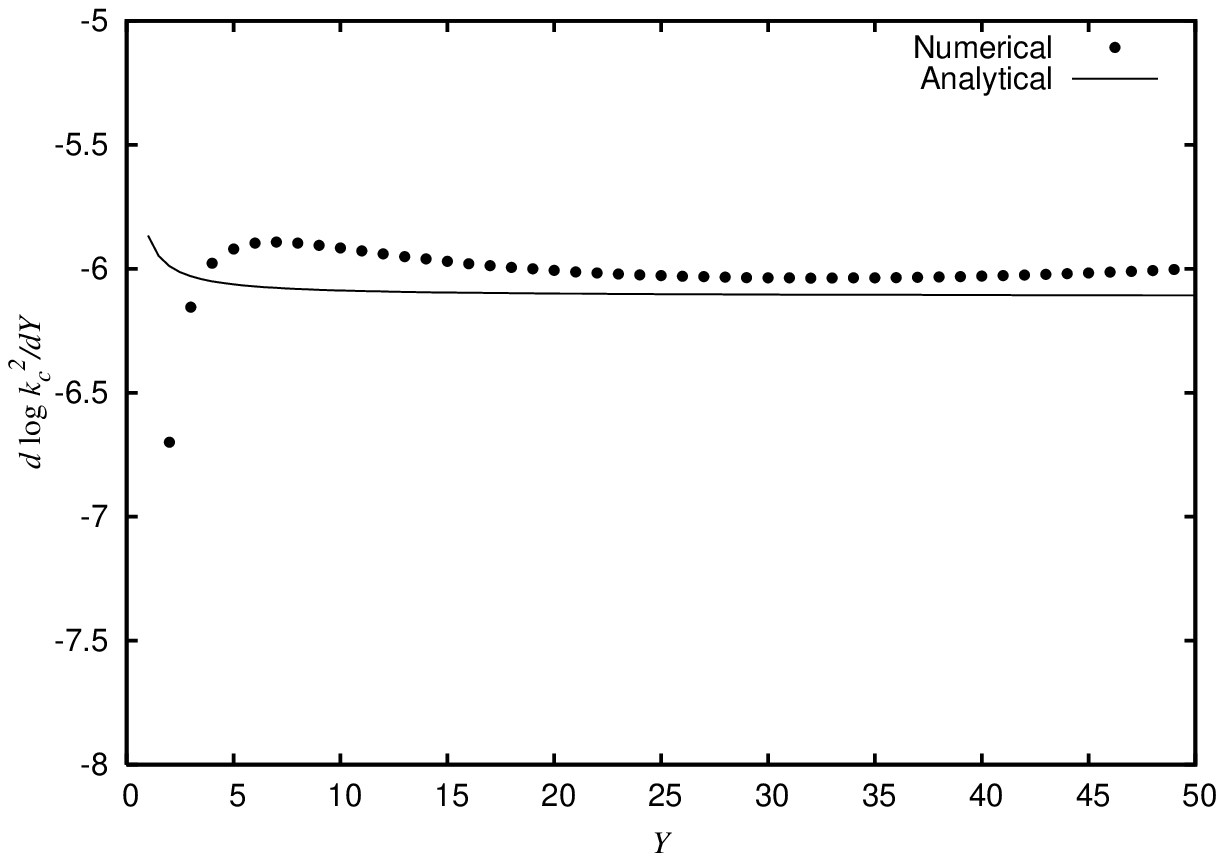,width=0.5\columnwidth}%
\epsfig{file=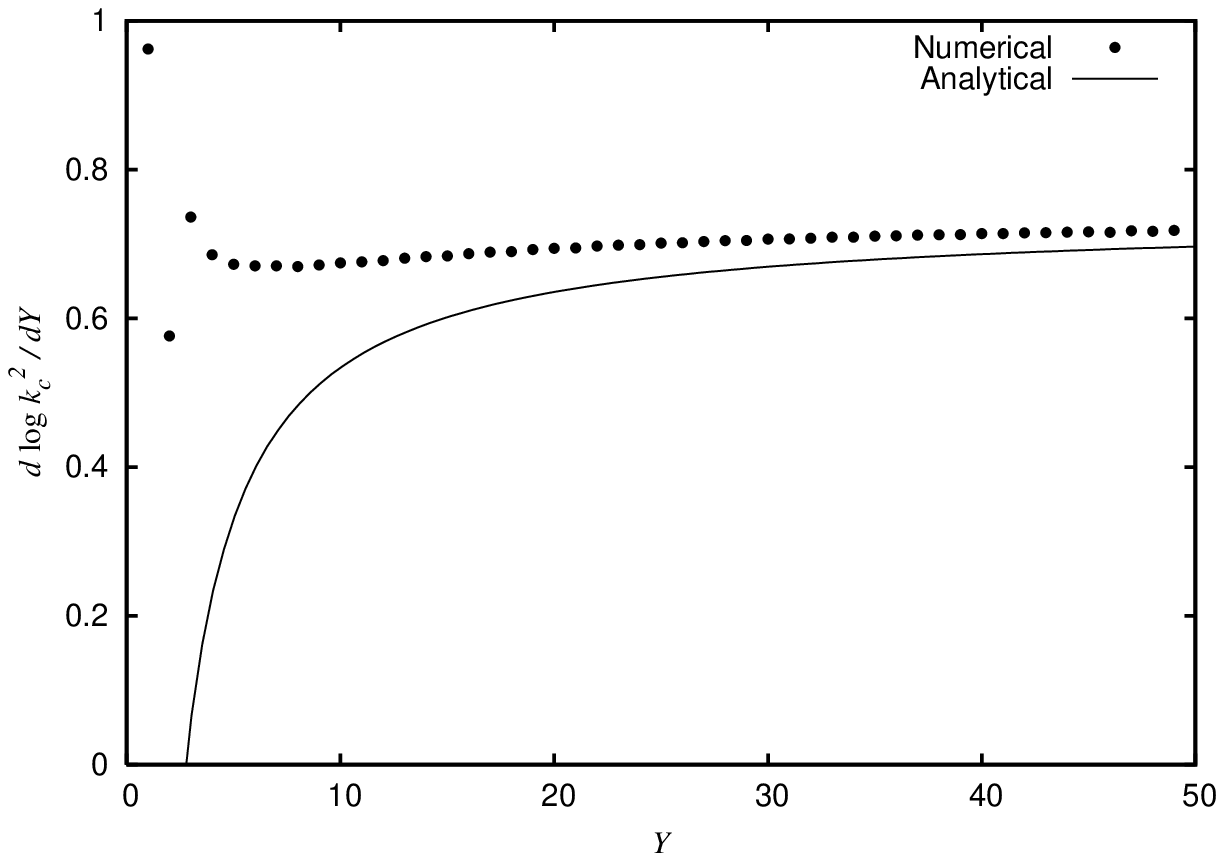,width=0.5\columnwidth}
\caption{Comparison of the analytical form for the derivative in 
\protect\eq{dqs} with the 
results from the numerical simulation, for backward evolution (left) and 
forward 
evolution 
(right). The equation analyzed here is given by the 
the two-pole toy kernel \protect\eq{twopole} with $\gamma_0=0.4$. Note the 
order 
of magnitude  difference between the backward (left) and forward (right) 
propagation speeds.}
\l{derivative-fig}
\end{figure}

The saturation scale $k_c(Y)$ is defined as the momentum, at a certain 
rapidity, 
where the 
amplitude reaches a particular value $\kappa$, i.e., 
${\cal T}(k_c(Y),Y)=\kappa$. The overall normalization therefore depends on 
the 
chosen value 
of $\kappa$ and it is most convenient to study the logarithmic derivative of 
$k_c(Y)$, where 
the normalization cancels. Because the shape of the front is not completely 
fixed as it 
propagates in rapidity, the functional form of ${\partial \log 
k_c(Y)}/{\partial 
Y}$ will 
also depend on $\kappa$, but this dependence is a subleading effect in $Y$ and 
vanishes as 
$Y$ becomes large\footnote{See \cite{NumBK} for examples and a thorough 
discussion.}.
Here we choose $\kappa=0.01$.

The analytical result given by \eq{minimal} is
\begin{equation}
\frac {\partial \log k_c(Y)}{\partial Y} = v_+ -\frac{3}{2\gamma_c Y},
\label{dqs}
\end{equation}
where subleading ${\cal O}(Y^{-3/2})$ terms have been omitted for sake of 
simplicity. Fig.\ 
\ref{derivative-fig} 
shows the results of the comparison between the simulations and expression 
\eq{dqs} for the 
two-pole toy kernel \eq{twopole} with $\gamma_0=0.4$ (the constants are given 
by 
Eq.\ 
\eq{frontvelocity}:
$v_+  =   0.7372$,
$v_-  =  -6.112$,
$\gamma^c_+  =  0.9157$, and 
$\gamma^c_-  =-0.2202$). The results indicate that the two results tend to the 
same 
asymptotic values $v_\pm$ of the derivative, while there are still appreciable 
subasymptotic 
effects \cite{NumBK}. A study of subasymptotic effects and early scaling, 
which we postpone for future work, could take benefit from the recent work in Ref.\ 
\cite{Marquet:2005ic}.

%=======================================================================
\section{Conclusions and physical implications}\label{Conclusions}
%=======================================================================

Let us summarize the results obtained in our study of the gluon momentum 
propagation (or equivalently, dipole size 
evolution) during nonlinear
QCD evolution at high density/energy:

(i) The  propagation of gluon momenta towards the 
ultraviolet and infrared with energy correspond to two different 
``universality 
classes'' of solutions of  the non-linear Balitsky--Kovchegov evolution 
equation. They are selected by the existence of an ultraviolet 
(resp.\ infrared) 
cut-off in the initial conditions but largely independent of the cut-off profile, if 
sharp enough.

(ii) The propagation  towards the ultraviolet is characterized by  asymptotic 
traveling 
wave solutions which are  independent of the (sharp enough) initial distribution. In 
particular, they are the same as those exhibited by  the onium-target amplitude with 
initial conditions defined by the QCD transparency property.
Hence,   the rate of evolution towards the ultraviolet is universally governed  by 
the 
saturation scale, generalizing the behavior of the BK dipole-target amplitude.

(iii) The   propagation of sizes towards the infrared shows new and distinctive 
features which have not been discussed, to our knowledge, in the literature 
previously. In the case of the LL BFKL kernel, i.e.\   in the leading log 
approximation of perturbative QCD, one obtains a scaling distribution (after a 
short diffusive interlude) which extends almost immediately  towards 
the far infrared region, revealing an infrared instability. This phenomenon is 
due 
to the singularity of the BFKL kernel 
at $\g =0.$ 

(iv) As soon as this singularity is shifted or smoothed out, new traveling wave 
solutions appear, with a (large) speed of backward propagation towards the 
infrared. In particular, next-to-leading corrections to the BFKL kernel  should 
give rise to  such a phenomenon and determine the value of the critical speed, 
as 
tested in different examples.

(v) The analytical predictions of the traveling wave pattern in both directions 
(ultraviolet and infrared) have been verified by numerical solutions of the 
evolution equations.

Our findings of the new traveling wave pattern towards the infrared within
a purely perturbative evolution framework could have sizable physical
implications. An example we have in mind is the problem of thermalization
of a Color Glass Condensate formed in the first stages of a 
heavy-ion collision.

Indeed, it is well known that it is not easy to reconcile the formation of a 
perturbative CGC (at weak coupling) as an initial compound state in a heavy
ion collision, with the hydrodynamic features of the final state which probably
require a fast thermalization or, at least, a rapid decrease of the mean free
path in the medium. We think that a discussion of this mechanism would be
interesting for the discussion of thermalization of an initial Color Glass
Condensate phase during a heavy ion collision process.

As a further outlook, it will also be worth examining the effect of the 
growth of the coupling constant during the propagation towards the infrared. 
Our present results have been obtained for simplicity assuming a fixed value 
of the coupling constant. They will have a natural extension to running 
coupling, using the method of Ref.~\cite{Munier} for solutions of the BK 
equation with running coupling. We expect qualitatively similar results but 
the quantitative analysis deserves to be performed.

These essentially geometric features  may also be modified by a better 
description of the interaction with the target. 
It is nowadays emphasized \cite{fluctuations,NumBK} that  fluctuations could  
modify the saturation solutions, violate geometric scaling or even exert a 
back-reaction 
onto 
the linear regime. What we obtain from our analysis of the ``universality 
class'' 
of propagation towards the ultraviolet will probably survive the implication of 
fluctuations. Even more interesting is the investigation of fluctuation effects 
on 
the infrared propagation. This will be the subject of forthcoming studies. 
Properties borrowed  from statistical physics could   help in attacking this 
problem with new tools.

\begin{acknowledgments}
We thank S.N. Majumdar for an inspiring seminar on statistical physics and A.\ 
Bialas and K.\ Golec-Biernat for fruitful 
discussions in a first stage of this work. C.\ Marquet and G.\ Soyez  are to be 
thanked for their clever remarks and careful reading of the manuscript. R.E.\ thanks 
the Service de Physique Th{\'e}orique of CEA/Saclay for hospitality when parts of 
this work was done.
\end{acknowledgments}

\end{document}